\documentclass[aps,preprint,floatfix]{revtex4}
\usepackage{epsfig}
\newcommand{\be}{\begin{equation}}
\newcommand{\ee}{\end{equation}}
\begin{document}

\title{Large $N$ QCD on the lattice -- A review of recent results in the
fermionic sector}

\author{R. Narayanan}
\affiliation{Department of Physics, \\
Florida International University, \\ 
Miami, FL 33199, USA\\ 
E-mail: narayanr@fiu.edu}

\author{H. Neuberger}
\affiliation{Department of Physics and Astronomy, \\ 
Rutgers University, \\
Piscataway, NJ 08855, USA\\
E-mail: neuberg@physics.rutgers.edu}  

\begin{abstract}
It is possible to numerically solve QCD in the planar limit using
standard numerical techniques on existing computer clusters.
The basic ideas behind the computational strategy and
recent numerical results in the fermionic sector of large $N$
QCD are reviewed.
\end{abstract}

\maketitle

\section{Introduction}

The large $N$ limit of QCD has several attractive features:
\begin{itemize}
\begin{item}
Only planar diagrams contribute
at fixed $g^2N$~\cite{thooft1}.
\end{item}
\begin{item}
Fermions in the fundamental representation are automatically quenched
as long as the number of flavors is kept finite when taking
$N\rightarrow\infty$~\cite{thooft1}.
\end{item}
\begin{item}
It is expected to preserve the essential perturbative and non-perturbative
features of the $N=3$ theory. The exact solution of the two dimensional theory 
reveals a tower of mesons~\cite{thooft2}.
\end{item}
\begin{item}
Phenomenological predictions from large $N$ QCD tend to be close to
experimental results indicating that leading corrections in the
$1/N$ expansion are at least occasionally small~\cite{manohar}.
\end{item} 
\begin{item}
Recent work indicates a duality between large $N$ gauge
theory and some string theory at zero string coupling~\cite{string}.
\end{item}
\end{itemize}

Lattice QCD provides a non-perturbative method to numerically solve
large $N$ QCD. Conventional methods involve computing physical
quantities as a function of $N$ and extrapolating these results to
the $N\rightarrow\infty$ limit~\cite{teper}.  So far, this approach
has only been used to compute observables in the pure gauge theory,
but is has provided evidence that large $N$ can be quantitatively
useful even at $N=3$.
Conventional computations proceed as follows: One
first takes the large volume limit followed by the continuum limit at a fixed
$N$ and then extrapolates to $N\rightarrow\infty$. This approach
will be difficult for fermionic quantities. This is due to
the fact that the quenched theory at finite $N$ has pathological
singularities~\cite{sharpe1}. Therefore, it is necessary to first
compute fermionic observables in a full dynamical simulation at
a fixed $N$ and then take the $N\rightarrow\infty$ limit.

A short cut to obtain results in the
$N\rightarrow\infty$ limit using lattice QCD
should use two aspects of large $N$
QCD. One is that the physical volume corrections are expected to be
small in the $N\rightarrow\infty$ limit since one has infinite number
of degrees of freedom at a single lattice site. Secondly, fermions are
naturally quenched in the $N\rightarrow\infty$ limit. To make use of
the second feature, it is best to take the $N\rightarrow\infty$ limit
of a quenched theory at a fixed lattice coupling, $b=1/(g^2N)$, before
taking the large volume limit and the continuum limit. Since we are in
a fixed lattice volume and lattice coupling, all fermionic observables
are well-defined and therefore the $N\rightarrow\infty$ limit will be
well-defined. Since we are taking the large volume limit in the
$N\rightarrow\infty$ theory, this is also well-defined and does not
suffer from quenched pathologies. Lastly, we are taking the
continuum limit of a set of well-defined theories on the lattice.

The procedure described in the previous paragraph will work independent
of the size of the finite volume corrections in the $N\rightarrow\infty$ limit.
But, as mentioned before, this correction is expected to be small.
In fact, tremendous excitement was generated by a paper of Eguchi and
Kawai~\cite{EK} in the early 1980s since it showed that the lattice
theory in the $N\rightarrow\infty$ limit reduced to a single site model
if the $U^d(1)$ symmetries specific to this lattice model remain
unbroken. This theory has only $d$ SU(N) matrices with $N\rightarrow\infty$
and the $U^d(1)$ symmetries are the $N\rightarrow\infty$ limit of
the $Z^d_N$ symmetries associated with the $d$ Polyakov loops in the
single site model. Unfortunately, the $U^d(1)$ symmetries are broken
in the weak coupling phase for $d>2$~\cite{BHN}. Since the reduction to
a single site model is very attractive, solutions to fix the above
problem were immediately found. One is referred to as the quenched
Eguchi-Kawai model~\cite{BHN} and another is referred to as the
twisted Eguchi-Kawai model~\cite{twist}. The twisted Eguchi-Kawai
model is an elegant solution to the problem encountered by the
original proposal of Eguchi and Kawai and has also led to the
discovery of non-commutative gauge field theories~\cite{noncomm}.
But, it is possible to include fermions only under some very
specific conditions in the twisted Eguchi-Kawai model.

Quenched Eguchi-Kawai solves the problem of the original Eguchi-Kawai
in an intuitive fashion and has other implications. The idea is to 
restrict the integral over all SU(N) matrices 
with a fixed set of $N$ eigenvalues. Observables
are first computed under this restriction and these observables are
then averaged over all choices for the $N$ eigenvalues. 
Freezing the eigenvalues of the SU(N) matrix in the computation of
physical observables can be interpreted as quenched momentum
prescription for gluons shown to be valid in the realm of
perturbation theory~\cite{Grokit}. In fact, the interpretation of
quenched momentum prescription extends to fermions~\cite{Levneu}. Fermion propagators
are computed in the pure gauge theory owing to the fact that fermion
loops are quenched in the $N\rightarrow\infty$ limit. But, fermion
propagators with momenta that differ by $p_\mu$ in a fixed
gauge field background are computed by using two different gauge
field backgrounds that differ by a $U(1)$ phase of $e^{ip_\mu}$ in
the $d$ directions. This is referred to as the quenched momentum
prescription for fermions. Since we integrate over all SU(N) fields,
momentum for a single propagator has no meaning. Only physical
observables like meson propagators carry momentum where the two
quarks that make up the meson see gauge fields that differ by
the U(1) phase. Furthermore, one sees that mesons can have
momentum as small as $2\pi/N$ on a single site lattice and therefore
one can study mesons for a continuum of momenta on a single site
lattice in the $N\rightarrow\infty$ limit. 

The proposal to numerically solve QCD in the large $N$ limit should
therefore proceed as follows. Consider the single site SU(N) lattice
gauge theory with the Wilson plaquette action. This model has $d$
SU(N) matrices denoted by $U_\mu$, $\mu=1,\cdots,d$.  Generate gauge
fields $U$ according to the single site Wilson plaquette action with
the added restriction that the $d$ SU(N) integrals go over all SU(N)
matrices with a fixed set of $N$ eigenvalues that are uniformly
distributed on the unit circle.  Compute quark propagators in the
fixed gauge field background $e^{ip_\mu} U_\mu$ where $p_\mu$ is
restricted to take values equal to $\frac{2\pi n_\mu}{N}$ such that
modified gauge field seen by the quark belongs to SU(N). This quark
propagator, $G(p,U)$ carries a momentum equal to $p_\mu$. A meson
propagator is computed using ${\rm Tr} G(p,U)G(q,U)$ and the result is
a meson carrying a momentum $P=p-q$. Averaging over all gauge field
configurations results in the meson propagator in the large $N$ limit
of QCD. To recover ``translational invariance'', it is necessary to
sum over all $p$ and $q$ for a fixed $P=p-q$.  This amounts to a
computation of $N^d$ propagators in a fixed gauge field background and
this is a time consuming numerical computation for $d=4$. But the idea
of momentum quenching can be tested in $d=2$ and the
results~\cite{knn1} agree with the exact solution of 't
Hooft~\cite{thooft2}.

Eigenvalues of $U_\mu$ are uniformly distributed on the unit circle
in the quenched Eguchi-Kawai model. Then, $e^{ip_\mu}U_\mu$ with
$p_\mu=\frac{2\pi n_\mu}{N}$ is just a cyclic permutation of all
the eigenvalues of $U_\mu$ and should be included in the average over
all gauge fields. But the numerical algorithm that generates the gauge
fields does not succeed in doing this and it is therefore necessary to
sum over all $p$ and $q$ for a fixed $P=p-q$ when computing meson
propagators. Since
it is necessary to take the large $N$ limit first, the
proposal in the previous paragraph is quite a bit more numerical involved
than one naively envisions. 
Therefore, it calls for a different realization of
the original Eguchi-Kawai idea. 
The reduction
argument of Eguchi and Kawai also applies to the lattice model on an
$L^4$ periodic lattice. Reduction will go through and there will not
be any finite volume effects if the $d$ $U(1)$ symmetries associated
with the $d$ Polyakov loops are not broken. Indeed, numerical analysis
shows that the above symmetries are not broken as long as $L > L_c(b)$
for $d=3$~\cite{nn1} and $d=4$~\cite{knn2}.
Furthermore, $L_c(b)$ scales properly
as one takes the continuum limit indicating the existence of a
physical scale $l_c$ such that large $N$ QCD is properly reproduced,
without finite volume effects if $l>l_c$~\cite{knn2}. This is referred
to as continuum reduction and it results in
a modification of the proposal in the previous paragraph. One works
on an $L^d$ lattice with $L$ just above $L_c(b)$ instead of working
on a $1^d$ lattice. The gauge fields are generated using the
standard single plaquette Wilson action with no restriction on the
gauge fields themselves. All other aspects of the proposal remains
the same, with the fermion momenta $p_\mu$ now going in steps of
$\frac{2\pi}{NL}$. This modified proposal has been shown to work
in practice resulting in the computation in the chiral condensate
in the large $N$ limit of QCD~\cite{nn2}. Currently, work is close
to completion in the computation of the pion decay constant, $F_\pi$,
in the large limit of QCD~\cite{nn3}.

\section{Overlap fermions coupled to large $N$ gauge theory}

Overlap fermions~\cite{overlap} preserve chiral properties of fermions
on the lattice and enables a study of fermionic observables in the
massless limit of quarks. This makes it possible to compute the low
energy parameters in the chiral Lagrangian of large $N$ QCD.
In addition, large $N$ gauge fields on the lattice
naturally come in disconnected subspaces
where each subspace is assigned a topological charge defined
using overlap fermions.

The main point is the presence of a lattice phase transition in the
large $N$ limit of pure gauge theory which is an extension of the
Gross-Witten phase transition~\cite{GW} in $d=2$. The single
plaquette Wilson action is given by
\be
S=bN\sum_{x,\mu\ne\nu} {\rm Re\ Tr} U_{\mu\nu}(x)
\ee
\be
U_{\mu\nu}(x) = U_\mu(x) U_\nu(x+\hat\mu) U^\dagger_\mu(x+\hat\nu)U^\dagger_\nu(x)
\ee
where $b=\frac{1}{g^2N}$ is kept fixed on the lattice as $N\rightarrow\infty$.
The continuum limit corresponds to $b\rightarrow\infty$. The lattice is
taken to be a periodic torus with $L$ sites in all directions. The link
variables, $U_\mu(x)$, and the plaquette variables, $U_{\mu\nu}(x)$, are
in SU(N).
Let $e^{i\theta_{\mu\nu}^k(x)};k=1,\cdots,N$ 
with $-\pi < \theta_{\mu\nu}^k(x) \le \pi$ denote the $N$ gauge invariant
eigenvalues of $U_{\mu\nu}(x)$. Consider the observable,
\be
p(\theta) = \frac{1}{Z}\int [dU_\mu(x)]
\frac{1}{NL^dd(d-1)}\sum_{x,\mu\ne\nu,k} \delta[\theta-\theta^k_{\mu\nu}(x)]
e^S;\ \ \ \  Z=\int [dU_\mu(x)] e^S,
\ee 
interpreted as the distribution of the plaquette eigenvalues.
This distribution should be uniform at $b=0$ and should peak around
$\theta=0$ in the weak coupling limit. In the
weak coupling limit at any finite $N$, one expects
a sharply peaked distribution at $\theta=0$ with $p(\pm\pi)$ being
highly suppressed. 
This becomes a phase transition in the $N\rightarrow\infty$
limit at $b=b_c^B$. The distribution is non-zero over the whole range of
$\theta$ if $ b < b_c^B$ and the distribution has a gap with
$p(\theta)=0$ for $|\theta| > \theta_c^B(b)$ if $b > b_c^B$.
The critical coupling is exactly at
$b_c(B)$ for large $L$ and this is a lattice transition with
no continuum counterpart.
The phase
transition is at
$b_c^B=0.5$ in $d=2$ for all $L\ge 1$
and is the Gross-Witten phase transition~\cite{GW}.
$b_c^B=0.4$ in $d=3$ for all $L\ge 3$ in $d=3$~\cite{nn1}
and
$b_c^B=0.36$ in $d=4$ for all $L\ge 9$ in $d=4$~\cite{knn2}.
It is a very strong first order phase transition in $d=4$
making it possible to stay in the metastable weak coupling side
of the phase for $5 \le L < 9$~\cite{knn2}.

Gauge fields on the weak coupling side of the above lattice phase
transition naturally fall into disconnected pieces. Since there
is an upper bound on the plaquette angle, $\theta$, there exist
two different gauge fields, $U$ and $V$ say, such that it is
not possible to continuously deform $U$ into $V$ without violating
the upper bound somewhere in-between. One can conjecture that
the disconnected pieces of the gauge field space as dictated
by the upper bound on the plaquette angle are in one-to-one
correspondence with the topological charge of the gauge field
defined using overlap fermions.

\begin{figure}[ht]
\epsfxsize = \textwidth
\centerline{\epsfbox{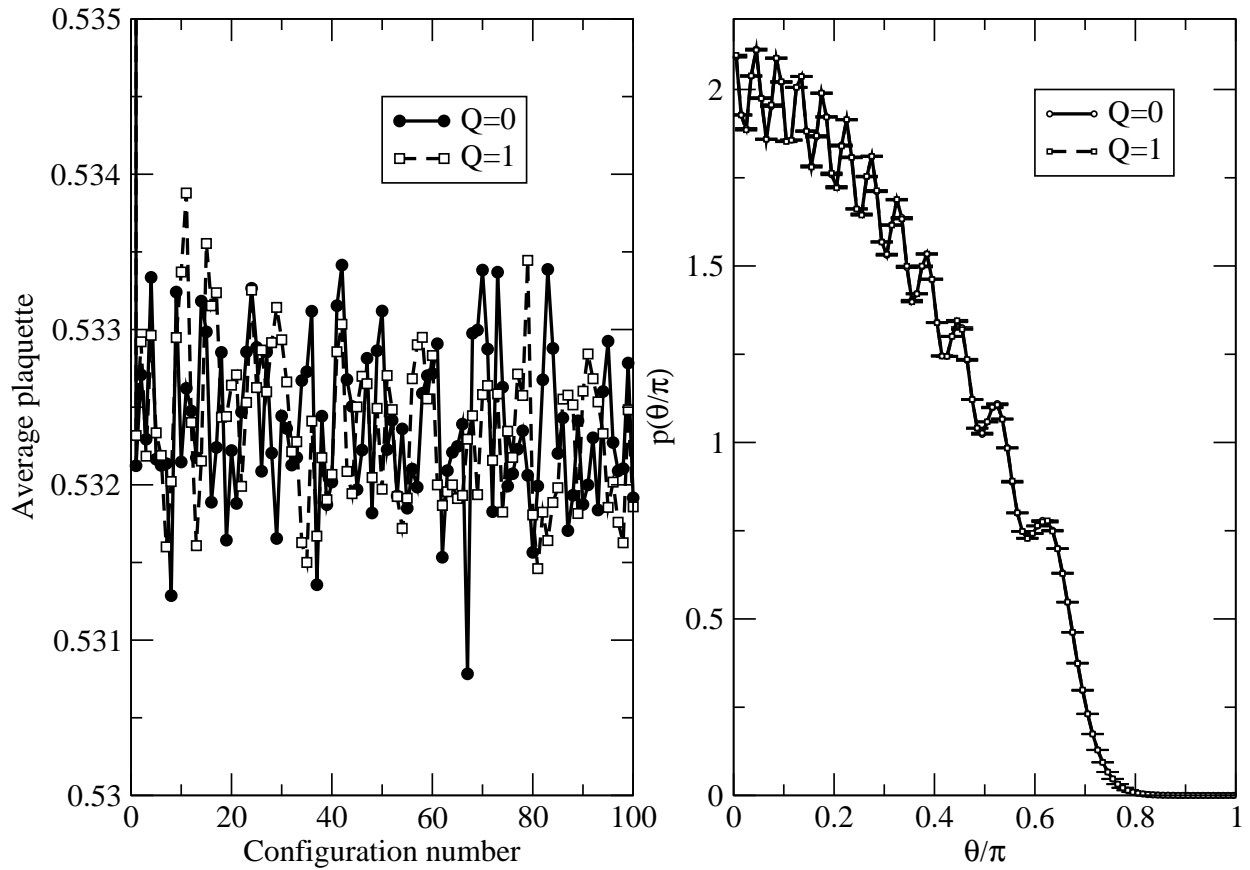}}
\caption{
Plaquette history (left panel) and plaquette eigenvalue distribution (right panel)
in the weak coupling phase. Two histories, one thermalized with $Q=0$ and
another with $Q=1$ are shown at $L=7$, $N=23$ and $b=0.35$.
\label{q01}}
\end{figure}

Consider a typical update of a large $N$ gauge field on the lattice
using a combination of Cabibo-Marinari heat-bath and SU(N)
over-relaxation~\cite{knn2} in the weak coupling phase.
A cold start where all gauge fields are set to unity will
result in thermalized gauge field configurations with zero
topological charge. There will be no tunneling events between
different topological charges
due to the gap in the eigenvalue distribution of the
single plaquette. One can obtain thermalized configurations
with different topological charges by starting from a gauge
field configuration with the appropriate topological charge
which also has a gap in the single plaquette eigenvalue distribution.
For example, one could start with a ``uniform'' instanton
configuration in $d=4$. The gauge field is abelian in nature
with a uniform field strength equal to $\frac{2\pi}{L^2}$
in a certain plane and a certain direction in color space~\cite{nn2}.
Fig.~\ref{q01} shows thermalized configurations in the
$Q=0$ and $Q=1$ sectors in the metastable weak coupling phase on a $7^4$ lattice
with $N=23$ at $b=0.35$. The two histories correspond to two different
starts where one is a cold start with $Q=0$ and another is a uniform
instanton start with $Q=1$. No tunneling was observed in either case.
The left panel shows the history of the
average plaquette and the right panel shows the distribution of the
plaquette eigenvalues. There is a gap in the distribution and both
of them look identical.
It is useful to note that a naive discretization of the classical
SU(2) instanton will not work since its plaquette eigenvalue distribution
will not satisfy the gap condition. In practice, the energy distributions
in the different topological sectors are very close to each other. Therefore,
it is possible to offer one thermalized configuration in one topological
sector as a global update while the gauge field is in a different 
topological sector and such an update has a chance of being accepted.

The gap in the eigenvalue distribution of the plaquette has implications
for the overlap Dirac operator in the weak coupling phase of large $N$ QCD.
The massless overlap
Dirac operator~\cite{overlap1} is defined as
\be
D_o = \frac{1+\gamma_5 {\rm sign}(H_w (M))}{2}
\ee
$H_w (M)$ is the Wilson Dirac operator at mass $M$, which should be chosen
with $-2 < M < 0$ and is an irrelevant parameter in the continuum limit
of the theory. Naively, one expects a gap in the spectrum of $H_w(M)$
around zero unless $M$ is close to the critical Wilson coupling and
by definition one needs to use a value of $M$ below this critical
coupling above. 
The flow of eigenvalues of $H_w(M)$ as a function of $M$ for a fixed
gauge field configuration is directly related to gauge field topology: 
The net number of levels that cross zero equals the topological charge~\cite{qtop}.
One can prove that a gap in the spectrum of
the eigenvalue distribution of the plaquette results in the gap
in the spectrum of $H_w(M)$ as long as one is significantly away from the
critical coupling~\cite{bound}. Unfortunately, the eigenvalue distribution
of the plaquette does not have a gap for $N=2$ or $N=3$ when gauge fields
are generated using the single plaquette Wilson action. As such, the spectrum
of $H_w(M)$ also does not have a gap below the critical coupling~\cite{wflow}.
But this is not the case in large $N$ QCD and this gives an unambiguous definition
of the topological charge. Furthermore, it also simplifies the numerical
evaluation of ${\rm sign}(H_w(M))$ which is the most time consuming part in
a computation involving the overlap Dirac operator.

\section{Phases of the large $N$ gauge theory}

As discussed in the previous section, one has to be in the weak
coupling phase of large $N$ QCD on the lattice, defined by the presence
of a gap in the eigenvalue distribution of the plaquette, in order to
properly reproduce continuum large $N$ QCD. The location of this lattice
transition is at $b=b_c^B$.
In addition, one has to
be in a phase where none of the $Z_N$ symmetries associated with
the Polyakov loops in the $d$ directions are broken.
We will refer to the strong coupling phase as $0$h since it is a ``hot''
phase with all $d$ $Z_N$ symmetries intact. On the weak coupling side,
there will in general be $d+1$ phases referred to as $0$c, $1$c, $\cdots$, $d$c
where $0,1,\cdots,d$ of the $d$ $Z_N$ symmetries are broken.
All the phase exist as long as $L\ge 3$ in d=3 and $L\ge 9$ in $d=4$.
Furthermore, one can be in a metastable $0$c phase for $5 \le L < 9$ since
the lattice transition separating the
$0$h and $0$c phase at $b=b_c^B$ is strongly first order.

\begin{figure}[ht]
\epsfxsize = \textwidth
\centerline{\epsfbox{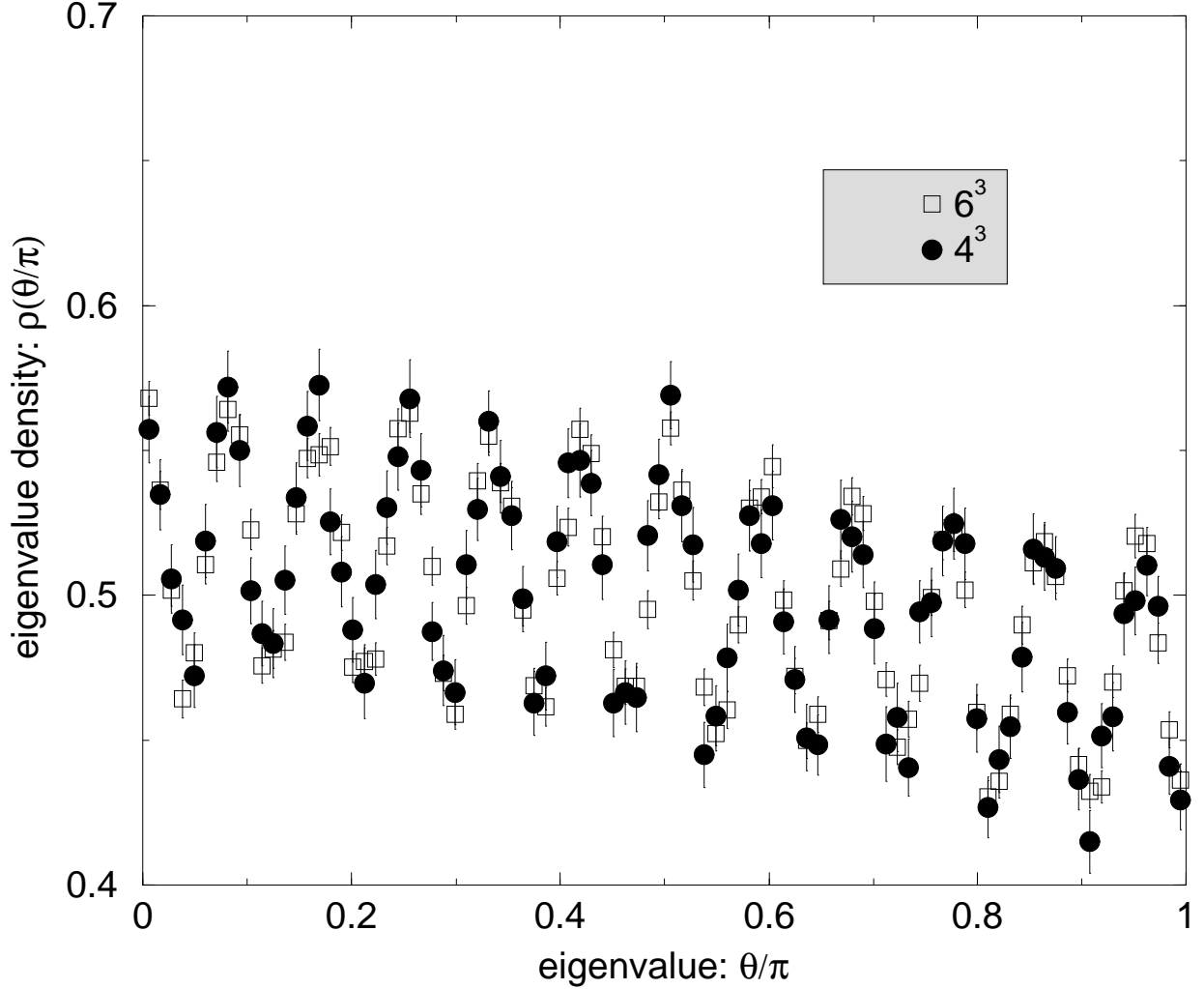}}
\caption{ Eigenvalue density distribution of a $4\times 4$ Wilson loop
on $4^3$(folded) and $6^3$(unfolded) at $b=0.66$ and $N=23$.
\label{fold}}
\end{figure}

\begin{figure}[ht]
\epsfxsize = \textwidth
\centerline{\epsfbox{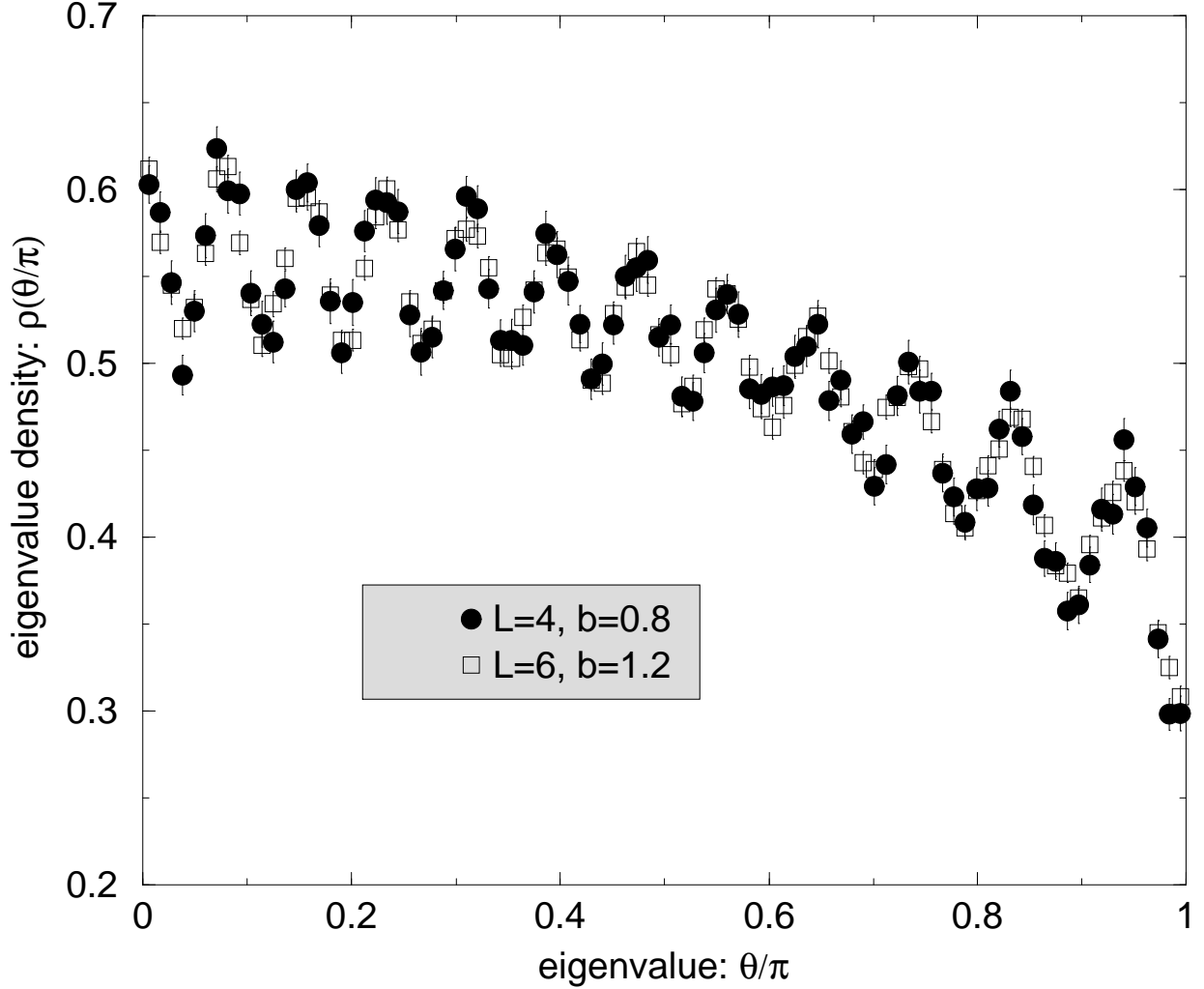}}
\caption{
Eigenvalue density distribution of $L\times L$ Wilson loop on
$L^3$ for $L/b=5$ and $L=4,6$. $N$ is set to $23$.
\label{scale}}
\end{figure}

Eguchi-Kawai reduction holds in the $0$c phase and large $N$ QCD is
properly reproduced in this phase without any finite volume effects.
In particular, consider a $K\times K$ Wilson loop on an $L^d$ 
periodic lattice with $K \le L$ or $K > L$. When $K > L$, the
Wilson loop is folded since some links appear more than once in
the operator. Reduction implies that the eigenvalue distribution of
all Wilson loops are independent of $L$. This is illustrated~\cite{nn1} 
in Fig.~\ref{fold} using
a $4\times 4$ Wilson loop on $4^3$ and $6^3$ lattice at a fixed
coupling and fixed $N$.
A proper realization of continuum large $N$ QCD also implies proper scaling
of observables in the $0$c phase and this is illustrated~\cite{nn1}
in Fig.~\ref{scale}
using a $L\times L$ Wilson loop on an $L^3$ lattice for fixed $L/b$ and
two different $L$ values.
`

\begin{figure}[ht]
\epsfxsize = \textwidth
\centerline{\epsfbox{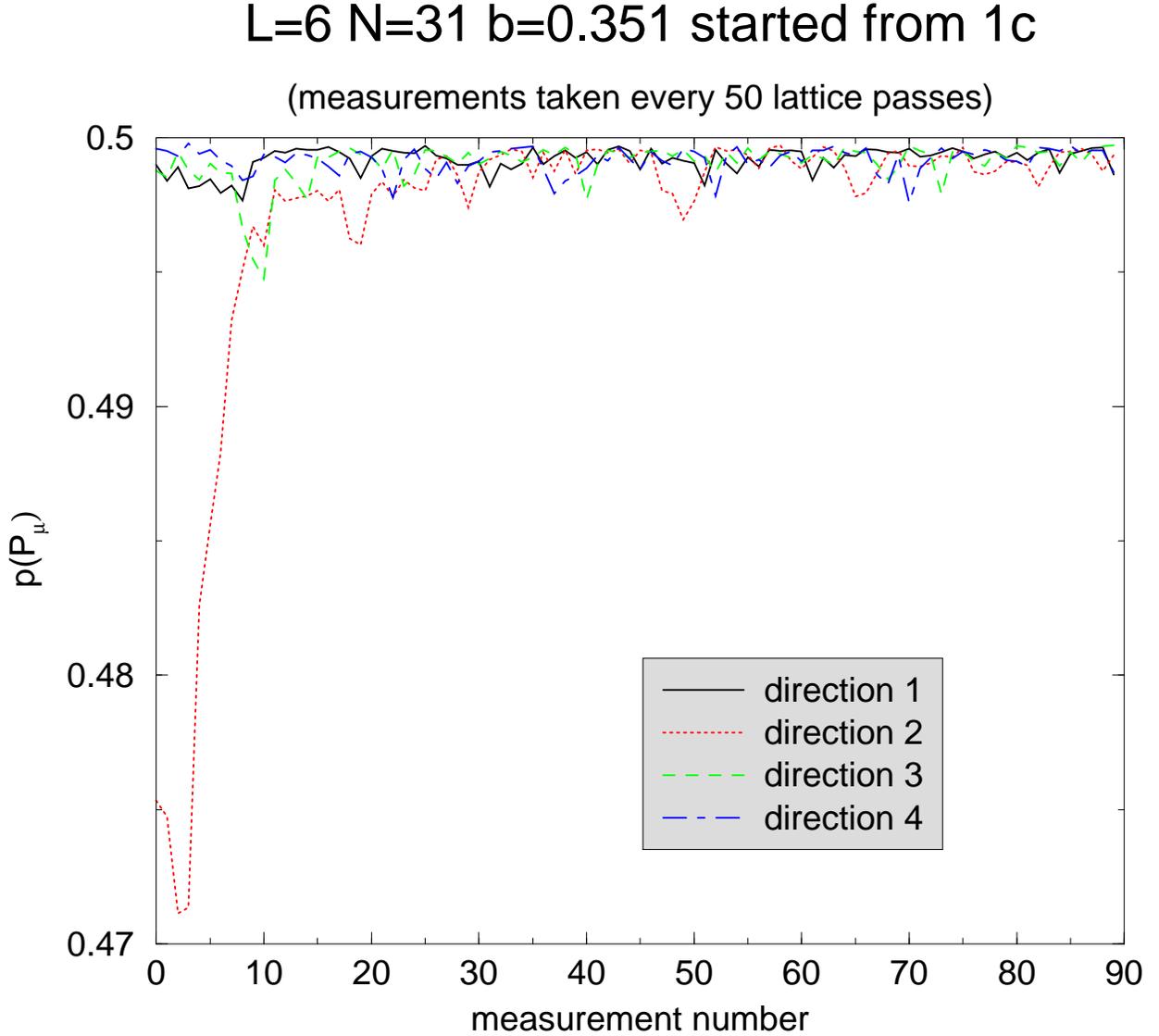}}
\caption{ History of the variable $p(\tilde P_\mu )$ for each direction.
We see the evolution from a state where one of the four $Z(N)$ 
factors is broken to one in which all four are preserved. 
\label{c1_to_0c}}
\end{figure}

\begin{figure}[ht]
\epsfxsize = \textwidth
\centerline{\epsfbox{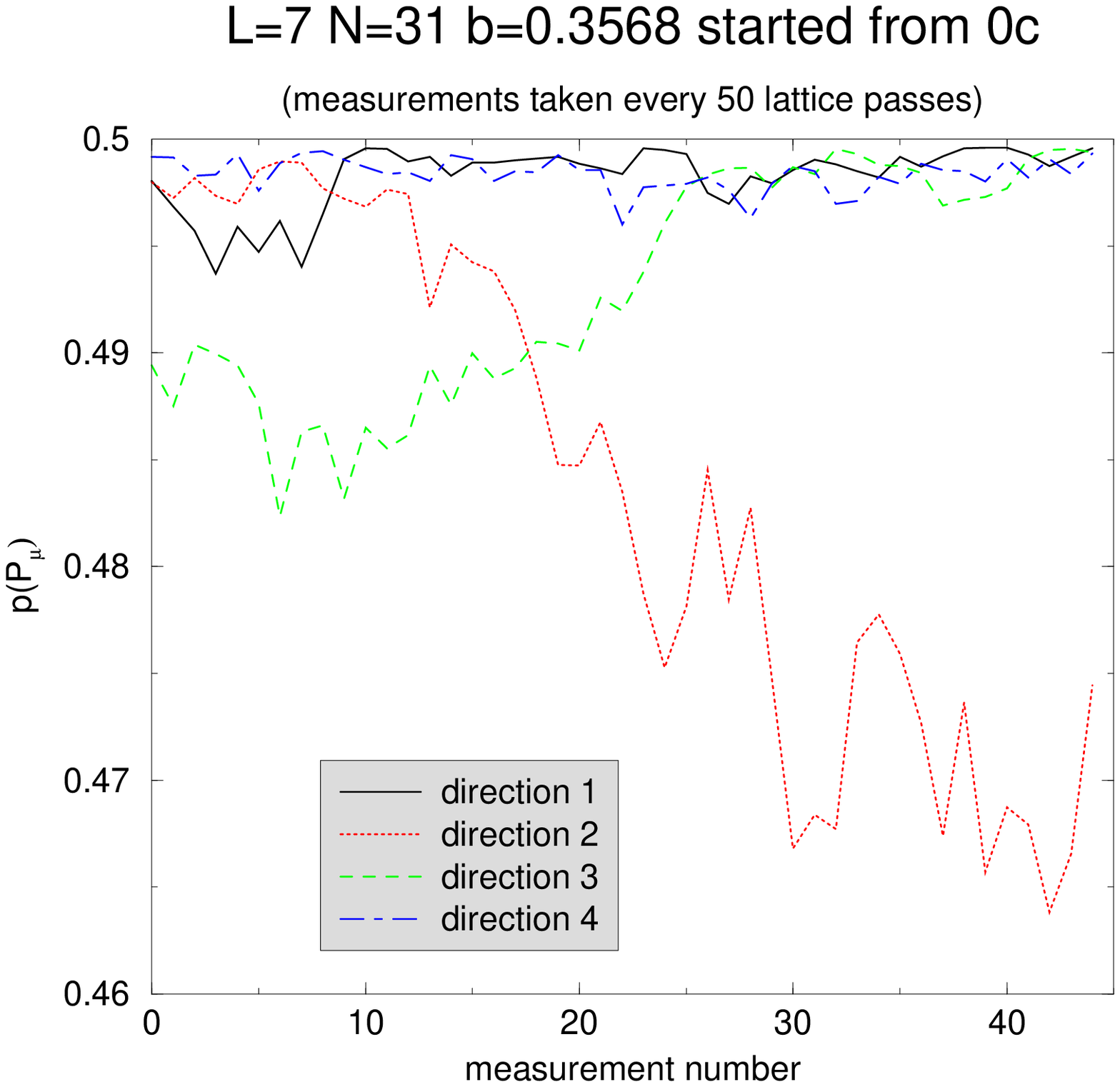}}
\caption{ History of the variable $p(\tilde P_\mu )$ for each direction.
We see the evolution from a state where all four $Z(N)$ factors are preserved
to one where one factor is broken. During the first fifty passes (before the
first measurement) Polyakov loops 
in direction 3 have acquired some structure 
but, ultimately, direction 2 is selected for breakdown and the Polyakov loops
in the other three directions converge to a symmetric state. 
\label{c0_to_1c}}
\end{figure}

Unlike the $0$h to $0$c transition, the rest of the transitions are
physical and therefore, the transition point will scale properly.
Our numerical analysis~\cite{nn1,knn2} indicates that all transitions
are distinct in the continuum limit but the main focus so far has
been on the $0$c to $1$c transition. 
One useful observable to locate the phase transition is~\cite{BHN} 
\begin{equation}
p(\tilde P_\mu) = \frac{1}{N^2} \langle \sum_{i,j=1}^N \sin^2 \frac{1}{2} 
(\theta^{\tilde P}_i - 
\theta^{\tilde P}_j )^2\rangle 
\end{equation}
where $\tilde P_\mu$ is the Polyakov loop (not its trace) in the
$\mu$ direction located at a fixed point in the 3-plane
perpendicular to $\mu$. The $e^{i\theta^{\tilde P}_j};\ \ j=1,\cdots,N$
are the $N$ eigenvalues of $\tilde P_\mu$.
The averaging is over all points in the 3-plane 
and over configurations. 
Equally spaced
eigenvalues respect the $Z(N)$ symmetry in the $\mu$ direction and 
maximize $p$ to 0.5. 
When
the eigenvalue spectrum starts getting modulated and opens a gap, 
$p$ drops below 0.5.
Fig.~\ref{c1_to_0c} shows that $b=0.351$ on $6^4$ lattice is in
the $0$c phase whereas Fig.~\ref{c0_to_1c} shows that $b=0.3568$
on $7^4$ lattice is in the $1$c phase.

\begin{figure}[ht]
\epsfxsize = \textwidth
\centerline{\epsfbox{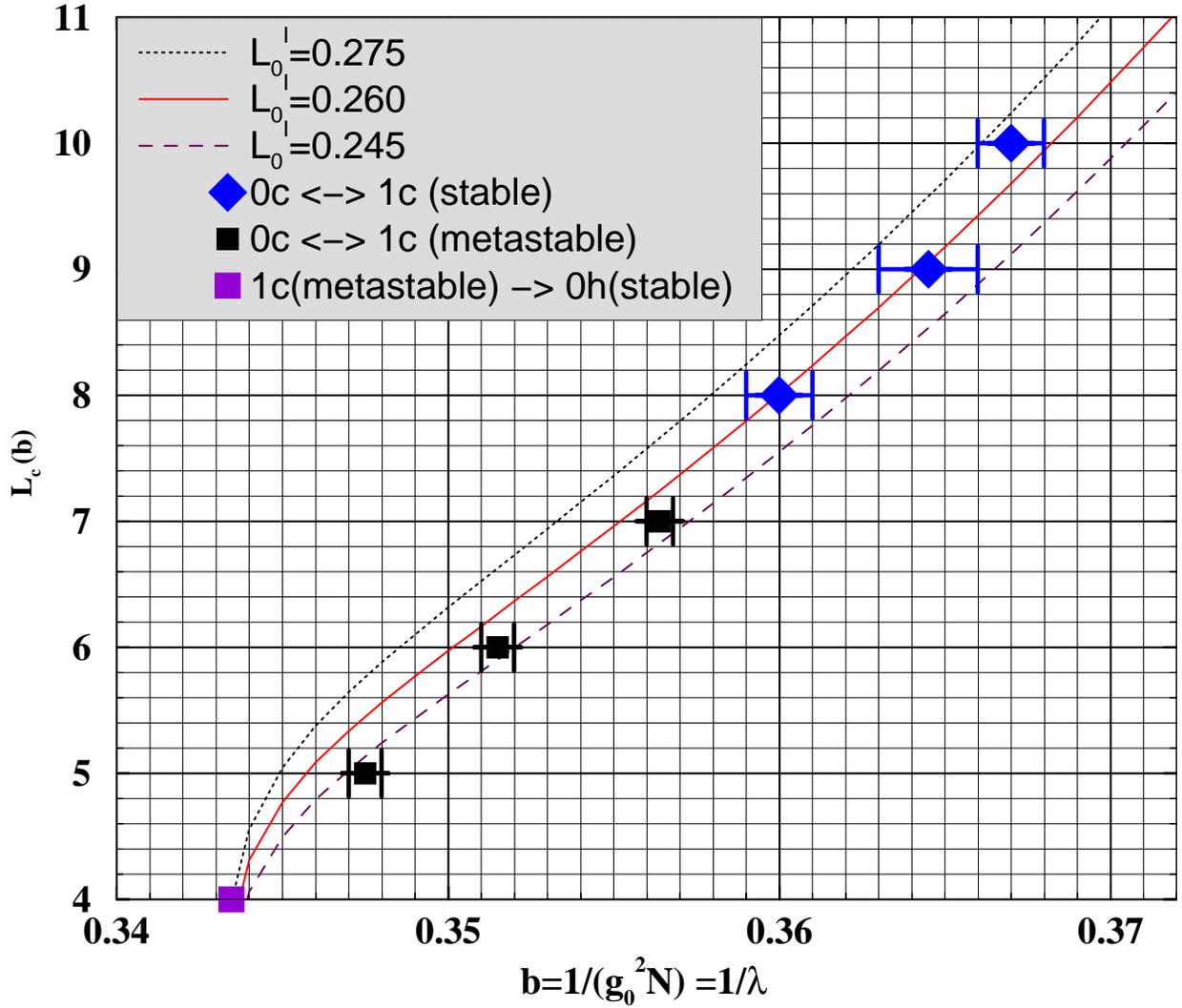}}
\caption{The transition ranges compared to possible two loop
renormalization group curves with tadpole improvement. 
\label{scomb}}
\end{figure}

Let $b_c$ be the location of the
numerically obtained transition on an $L^d$ lattice. This can be
inverted to define a critical size, $L_c(b)$,
that denotes the $0$c to $1$c transition. 
The coupling $b$ has
dimensions of length in d=3 and a numerical analysis~\cite{nn2} shows that
$L_c(b) = 4.6(4) b$. The 
numerically obtained~\cite{knn2} critical size $L_c(b)$ in $d=4$
shown in Fig.~\ref{scomb}
scales according to two loop tadpole improved~\cite{tadpole}
improved perturbation theory as
\be
L_c(b) = 0.260(15) \left( \frac{11}{48\pi^2b e(b)} \right)^\frac{51}{121}
\exp\left[ \frac{24\pi^2b e(b)}{11} \right]
\ee
with 
\be
e(b) = \frac{1}{N} \langle {\rm Tr} U_{\mu\nu}(x)\rangle
\ee

The $0$c phase of the large $N$ continuum theory on an $l^4$ torus is
clear and it describes large $N$ QCD at zero temperature.
The $4$c phase can
be interpreted as a continuum theory in finite volume and temperature
with temperature equal to $1/l$.  
Rotational invariance is
spontaneously broken in the $1$c, $2$c and $3$c phases of the
continuum theory. 
The physical interpretation of theses phases 
is not obvious.  
The $1$c quite possibly describes large $N$ QCD in the
deconfined phase since the finite size effects are only felt in the
one broken direction and this could be interpreted as finite
temperature. Then, one would conclude that large $N$ QCD in the
confined phase does not depend on temperature, a view that has been
recently discussed~\cite{cohen}.  The $3$c phase possibly describes a
continuum theory in a small finite box at zero temperature. The
continuum theories described in the phases other that $0$c and $1$c
might have implications not directly related to large $N$ QCD. A
careful study of the $0$c to $1$c transition would tell us if the
transition is the physical finite temperature phase transition. The
critical box size~\cite{knn2} associated with the $0$c to $1$c
transition in units of string tension is $l_c\sqrt\sigma \approx
1.56$.  A comparison with the numerical estimates of the critical
temperature~\cite{lucini} indicate that $l_c T_c=1$.  Recent numerical
studies indicate a finite latent heat associated with this
transition~\cite{kiskis} providing further supporting evidence for
this to be the physical finite temperature phase transition.

\section{Spontaneous chiral symmetry breaking}

Spontaneous chiral symmetry breaking is an important phenomenon in
QCD. The mass of the pion made up of two light quarks is not equal or
close to the sum of quark masses but proportional to the square root
of the quark mass since the pion is a pseudo-Goldstone boson
associated with spontaneous chiral symmetry breaking in QCD with
massless quarks. Several lattice studies have investigated this
phenomenon in the past using Wilson fermions~\cite{wilq} and staggered
fermions~\cite{stagq} and more recently using overlap
fermions~\cite{overq}. Quenched lattice studies of this phenomenon
suffer from quenched pathologies. The quenched pathologies arise out
of the fact that the singlet meson, $\eta^\prime$, becomes a
pseudo-Goldstone boson in the quenched
approximation~\cite{sharpe2}. This introduces a new mass parameter
$m_0$ in the chiral Lagrangian with $m_0 \approx 900$MeV in the chiral
limit. The pion mass no longer is linear in the square root of the
light quark mass and instead is given by~\cite{sharpe2} \be m_\pi^2
\propto m^{\frac{1}{1+\delta}};\ \ \ \ \delta=\frac{m_0^2}{N(4\pi
f_\pi)^2}\approx \frac{0.6}{N}.  \ee where $m$ is the light quark mass
and $f_\pi=93$MeV is the pion decay constant.  Spontaneous chiral
symmetry breaking occurs due to the presence of a non-zero density of
eigenvalues of the massless Dirac operator near zero.  Let
$\rho(\lambda)$ denote the density of the paired non-zero $\pm
i\lambda$ of the anti-hermitian massless Dirac operator after removing
the exact zero eigenvalues due to gauge field topology.  The chiral
condensate~\cite{banks}, \be \lim_{m\rightarrow 0}
\lim_{V\rightarrow\infty} \langle\bar\psi\psi(m)\rangle = \pi \rho(0),
\ee is non-zero if $\rho(0)$ is non-zero and this shows spontaneous
chiral symmetry breaking. The effect of the $m_0$ term in quenched QCD
results in a diverging chiral condensate~\cite{sharpe2}, given by \be
\lim_{V\rightarrow\infty} \langle\bar\psi\psi(m)\rangle \propto
m^{-\frac{\delta}{1+\delta}} \ee Typical quenched calculations on the
lattice focus on the behavior of pion mass as a function of quark mass
to extract the quenched divergence. Since $\delta > 0$, the effect of
quenched divergence is to make the pion mass at a fixed quark mass a
little heavier than what it should be in unquenched QCD. Since all
calculations are performed at finite volume, the effect due to
quenched divergence cannot be distinguished from finite volume
effects~\cite{columbia}. On the other hand, the chiral condensate
diverges in the quenched approximation and this effect goes in the
opposite direction from that of finite volume effects.  In the large $N$
limit, $\delta$ goes to zero and there are no quenched pathologies.

In order to avoid quenched pathologies, one has to take the large $N$ limit
before one takes the infinite volume limit. Furthermore, Eguchi-Kawai
reduction holds in the $0$c phase and therefore the infinite $N$ limit
in this phase will not depend on the physical volume. We should therefore
see evidence for a chiral condensate on a finite lattice in the large $N$
limit as long as we are in the $0$c phase. This is counter-intuitive since
we expect to see a finite density of eigenvalues near zero in the spectrum
of the massless Dirac operator even at finite lattice volume. Chiral
random matrix theory~\cite{rmt} provides an understanding of spontaneous chiral
symmetry breaking in large $N$ QCD on a finite lattice~\cite{nn2}. 

Simple counting of $N$ degrees of freedom shows that $\langle\bar\psi\psi\rangle$
is proportional to $N$. Therefore the low lying eigenvalues of the massless 
Dirac operator should scale like $\frac{1}{NV}$ if chiral symmetry is
spontaneously broken in the large $N$ limit. This is a consequence of level
repulsion. The paired non-zero eigenvalues, $\pm\lambda$, of the massless hermitian
overlap Dirac operator satisfies $0 < \lambda < 1$. The operator has
$2NV$ positive eigenvalues (assuming zero topology) on a $V=L^4$ lattice.
Level repulsion would result in eigenvalues that are roughly equally spaced.
Therefore, the mean spacing will be proportional to $\frac{1}{NV}$ and the
lowest eigenvalues will scale like $\frac{1}{NV}$. Furthermore, the scaled
eigenvalues $z=\lambda\Sigma NV$ should essentially be distributed in 
some universal manner and this universal behavior is dictated by chiral
random matrix theory~\cite{shuryak}. The scale, $\Sigma$, that connects
$z$ and $\lambda$ is the chiral condensate. Universal behavior is approached
as $NV$, the number of eigenvalues, go to infinity. This can be achieved
by keeping $N$ fixed and taking $V$ to infinity which is the conventional
approach. But, it can also be achieved by taking $N$ to infinity at fixed
$V$ and this makes use of the reduction in large $N$ QCD. 
Quenched divergences imply that $\lambda$ does not scale like
$\frac{1}{NV}$ at fixed $N$ as $V\rightarrow\infty$. In spite of this,
it has been shown that one can define a scale $\Sigma(V)$ such that
chiral random matrix theory is respected in the quenched approximation~\cite{daam}.
But, $\Sigma(V)$ diverges as $V\rightarrow\infty$.

One could show evidence for spontaneous
chiral symmetry breaking in large $N$ QCD by a direct computation of
$\Sigma$ in the $0$c phase. This can be done by a computation
of $\lim_{m\rightarrow 0}\lim_{N\rightarrow\infty}\langle\bar\psi\psi(m)\rangle$ 
on a finite $L^d$ lattice. In this case, it is necessary to take the large $N$
limit before taking the massless limit since the chiral condensate is zero
at $m=0$ for any finite $N$. Numerically, one has to perform some sort of
a finite $N$ analysis equivalent to the usual finite volume analysis
done in order to extract an order parameter in the thermodynamic limit.
Chiral random matrix theory provides us with a way to perform the
finite $N$ analysis without a computation of the chiral condensate.
Let $\lambda_i$, $i=1,\cdots,2NV$, with $\lambda_1 < \lambda_2 < \cdots \lambda_{2NV}$
be all the positive eigenvalues of the hermitian overlap Dirac operator.
Let $z_k=\lambda_k \Sigma N V$ be the scaled eigenvalues that obey the 
universal distribution given by chiral random matrix theory~\cite{daam2}.
As $N$ increases, more and more $z_k$ will obey the universal distributions.
In particular, one will find that $p(r=\frac{\lambda_1}{\lambda_2})$ obeys a universal
distribution as long as $N$ is large enough. Once this is achieved, 
$\Sigma$ can be obtained by matching $\langle z_1\rangle$ and
$\langle z_2\rangle$ to their values given by chiral random matrix theory.
A numerical computation shows that one can obtain $\Sigma$ in the large $N$
limit of QCD using chiral random matrix theory~\cite{nn2}. 
Existence of a chiral condensate in the continuum limit amounts to showing
that $\Sigma(b)$ obtained on the lattice scales properly in the continuum limit
and one finds that $\Sigma^{1/3} l_c = 0.588$ which translates to
$\Sigma^{1/3} \approx 155$MeV if we assume that $\frac{1}{l_c}=T_c=264$MeV.

\subsection{Chiral condensate in the large $N$ limit of QCD${}_2$}

\begin{figure}[ht]
\epsfxsize = \textwidth
\centerline{\epsfbox{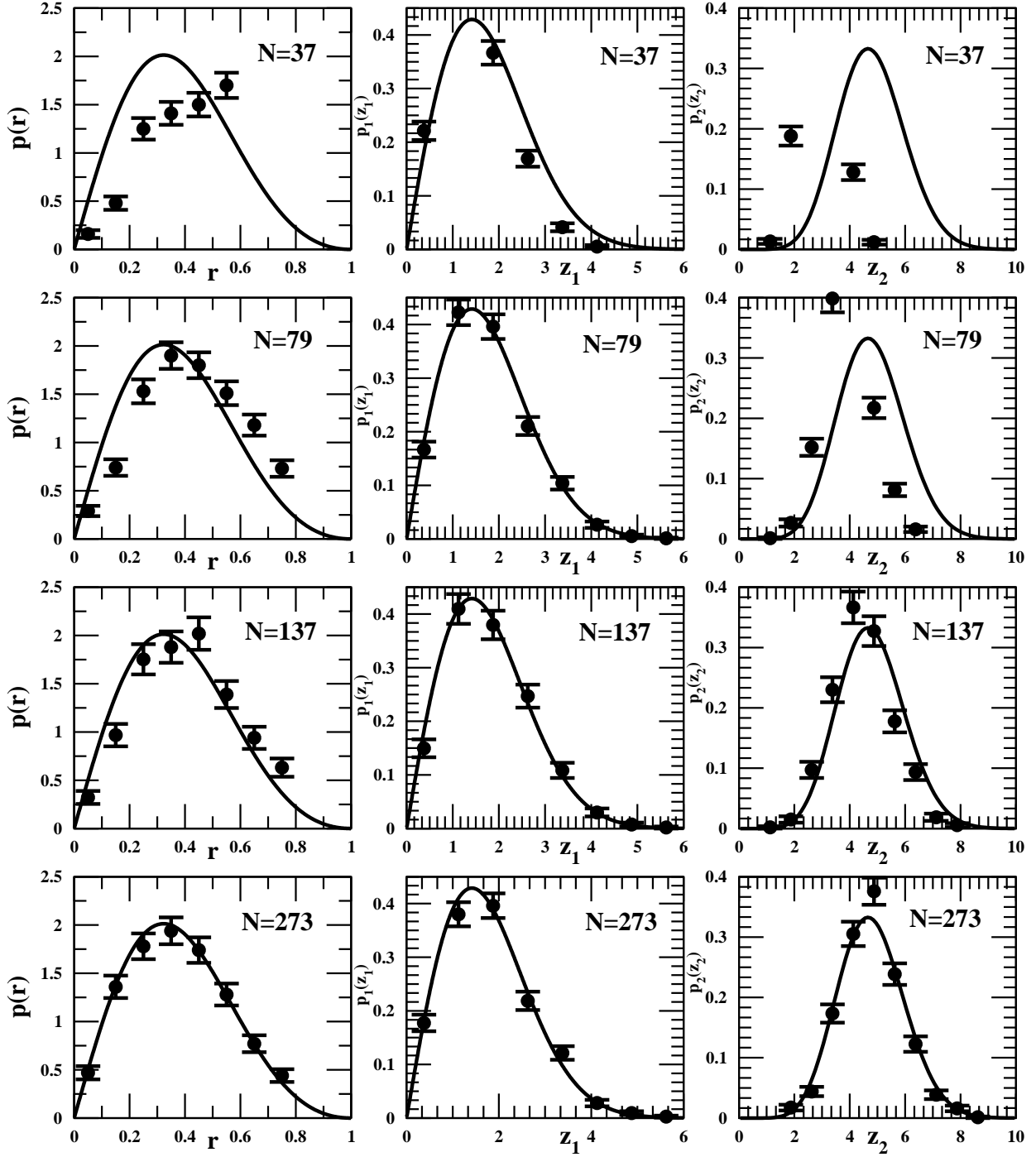}}
\caption{Approach to chiral random matrix theory on a $1^2$ lattice at
$b=1$ as $N$ is increased from $37$ to $273$.
\label{rmt2d}}
\end{figure}

\begin{figure}[ht]
\epsfxsize = \textwidth
\centerline{\epsfbox{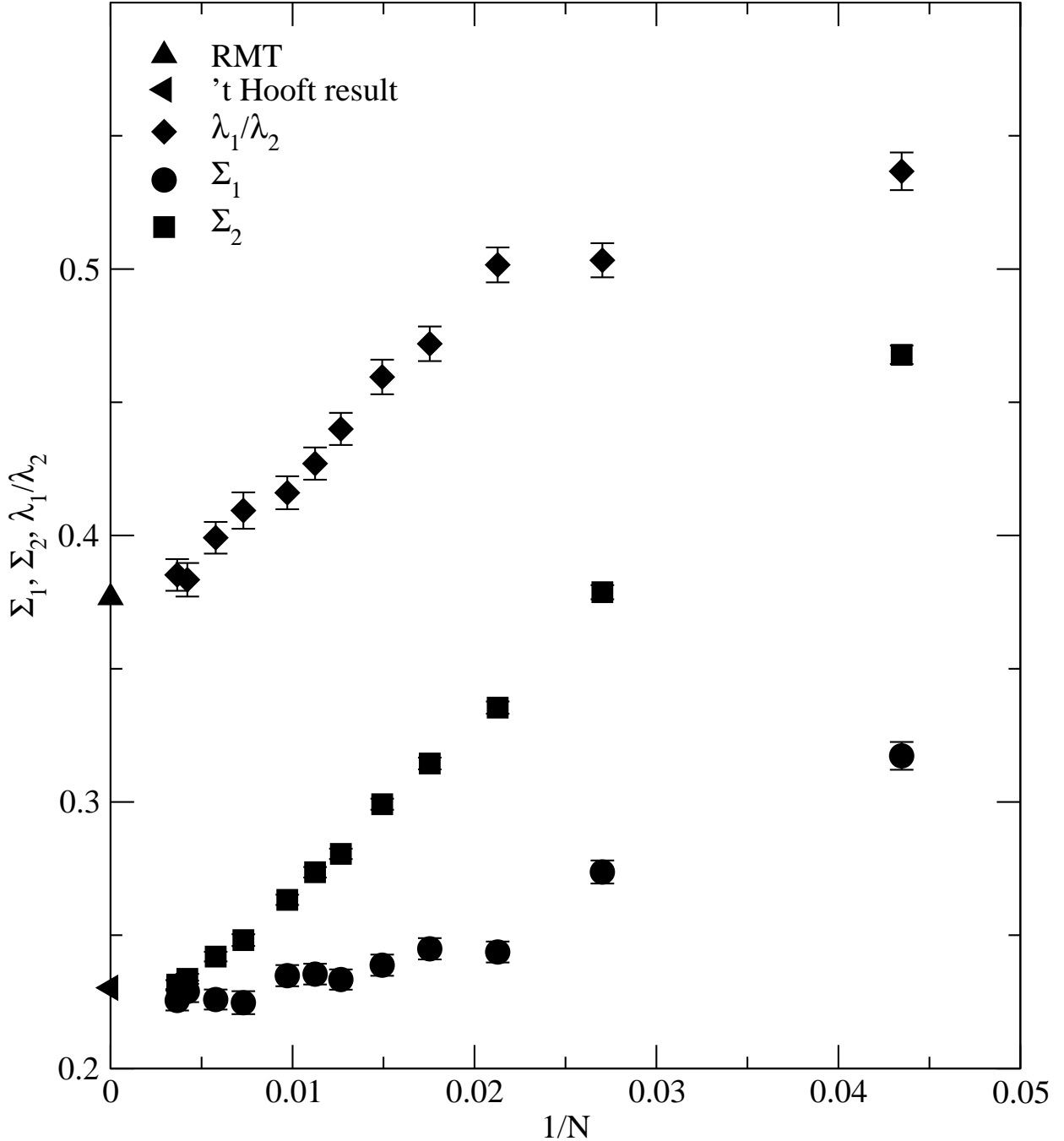}}
\caption{Estimation of the chiral condensate in the $N\rightarrow\infty$ limit
of QCD${}_2$ at
$b=1$ as $N$ is increased from $23$ to $273$.
\label{sigma2d}}
\end{figure}

Before proceeding to the computation of chiral condensate using chiral
random matrix theory, it is useful to verify the ideas of chiral random
matrix theory in two dimensional QCD~\cite{nn2}. The chiral condensate in the
't Hooft model is exactly known~\cite{zhit} and it is given by
\be
\Sigma(b) = \frac{|M|}{\sqrt{6\pi b}}
\ee
for the overlap Dirac operator,
where $M$ is the Wilson mass parameter used in the definition of the
overlap Dirac operator. Eguchi-Kawai reduction holds for all $L$ in $d=2$
and therefore one can compute the chiral condensate by taking the
large $N$ limit on a $1^2$ lattice. One can compute the full spectrum of
the overlap Dirac operator for very large $N$ using exact diagonalization
on a $1^2$ lattice. Let $p_1(z_1)$ and $p_2(z_2)$ be the universal distributions
of the two lowest eigenvalues and let $p(r)$ be the distribution of the ratio
of the first eigenvalue and the second eigenvalue. Fig.~\ref{rmt2d} shows that
approach to chiral random matrix theory as $N$ is increased on a $1^2$ lattice
at a fixed $b$. $N=37$ is too small for chiral random matrix theory to be
satisfied. At $N=79$ the first eigenvalue already satisfies the universal
distribution and both eigenvalues satisfy the universal distribution beyond
$N=137$. The coupling was set to $b=1$ and we used $M=-1$ in the definition
of the overlap Dirac operator. Therefore, we should find $\Sigma=0.23$
as $N\rightarrow\infty$. Furthermore, $<r>=0.37$ as $N\rightarrow\infty$.
The approach to the $N\rightarrow\infty$ limit of $<r>$ and $\Sigma$ using the
first and second eigenvalue is shown in Fig.~\ref{sigma2d}. One can see
that the first eigenvalue approaches chiral random matrix theory before the
second eigenvalue.

\subsection{Chiral condensate in the large $N$ limit of QCD}

Analysis of QCD${}_2$ shows that one can use chiral random matrix
theory to extract the chiral condensate in the large $N$ limit of
QCD. To do this, one has to be in the $0$c phase of QCD. Therefore, we
worked with $L$ in the range of $6$ to $10$. At each $L$, there is a
range of coupling $b_c^B < b < b_c(L)$ where the theory is in the $0$c
phase. It is best to pick the coupling close to $b_c(L)$ to minimize
lattice spacing effects at that $L$. We therefore used~\cite{knn2}
$b=0.350$ for $L \ge 6$, $b=0.355$ for for $L \ge 7$ and $b=0.3585$
for $L \ge 8$.  We worked with $L=6,7,8$ at $b=0.350$ to ensure that
reduction holds.  We worked with $L=8,9,10$ at $b=0.355$ again to
ensure that reduction holds.  We worked with $L=9$ at $b=0.3585$ and
we also obtained one result on a coarse lattice of $b=0.346$ at
$L=9$. 
The scaling properties of the code are as follows. The gauge field
generation scales like $N^3L^4$.
The
computation of the lowest two eigenvalues of the hermitian overlap
Dirac operator scales like $N^3L^8$ and this arises from two factors.
One factor is due to the action of the Wilson Dirac operator on a vector 
and this scales
as $N^2L^4$. The low lying eigenvalues of the overlap Dirac operator
are obtained using Ritz techniques and the relevant condition number
is the ratio of the highest to the lowest eigenvalue and therefore
the second factor scales
as $NL^4$. 

\begin{figure}[ht]
\epsfxsize = \textwidth
\centerline{\epsfbox{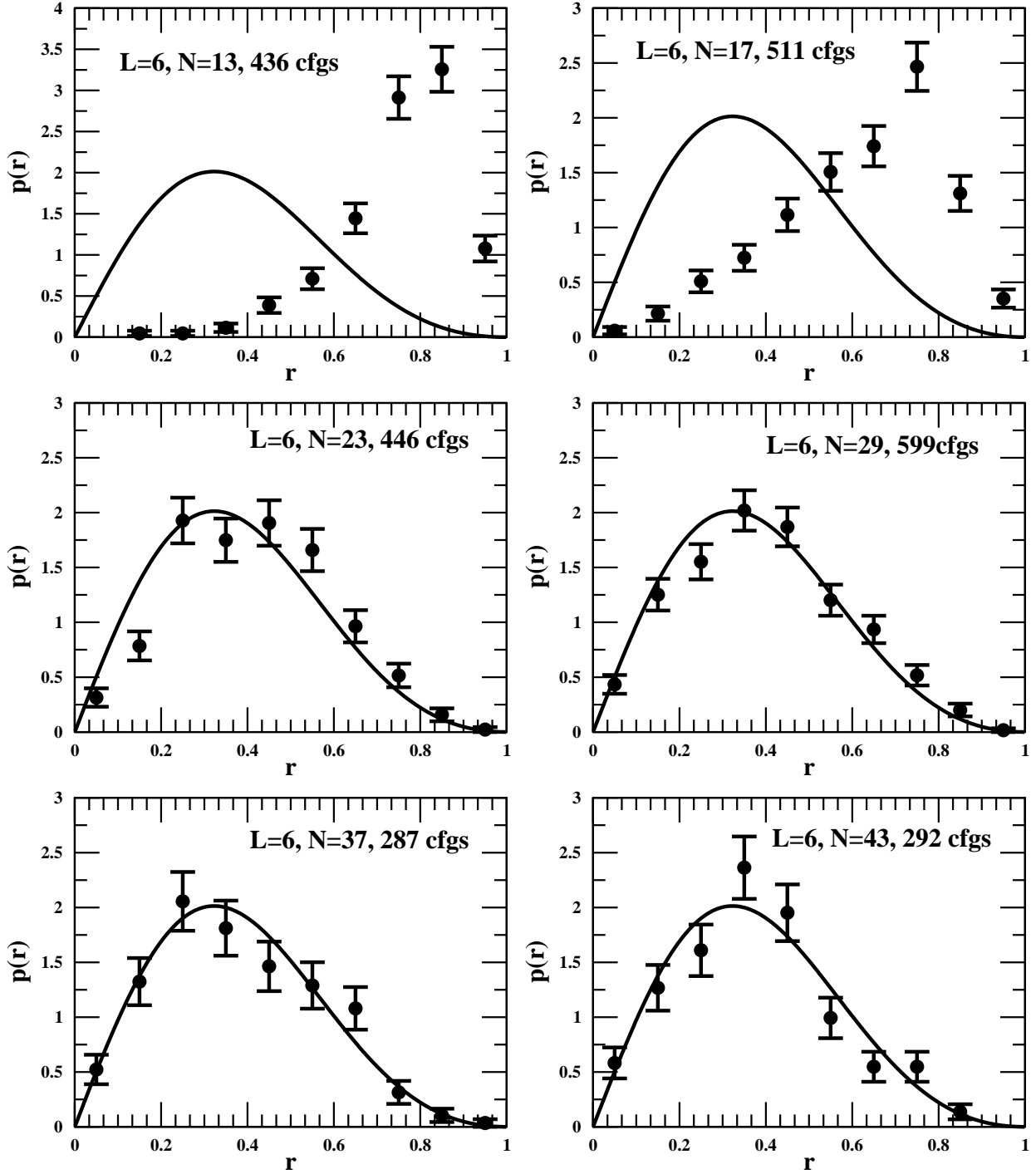}}
\caption{Approach to chiral random matrix theory at a fixed $b$ and $L$ 
with increasing $N$.
\label{d4rmt}}
\end{figure}

\begin{figure}[ht]
\epsfxsize = \textwidth
\centerline{\epsfbox{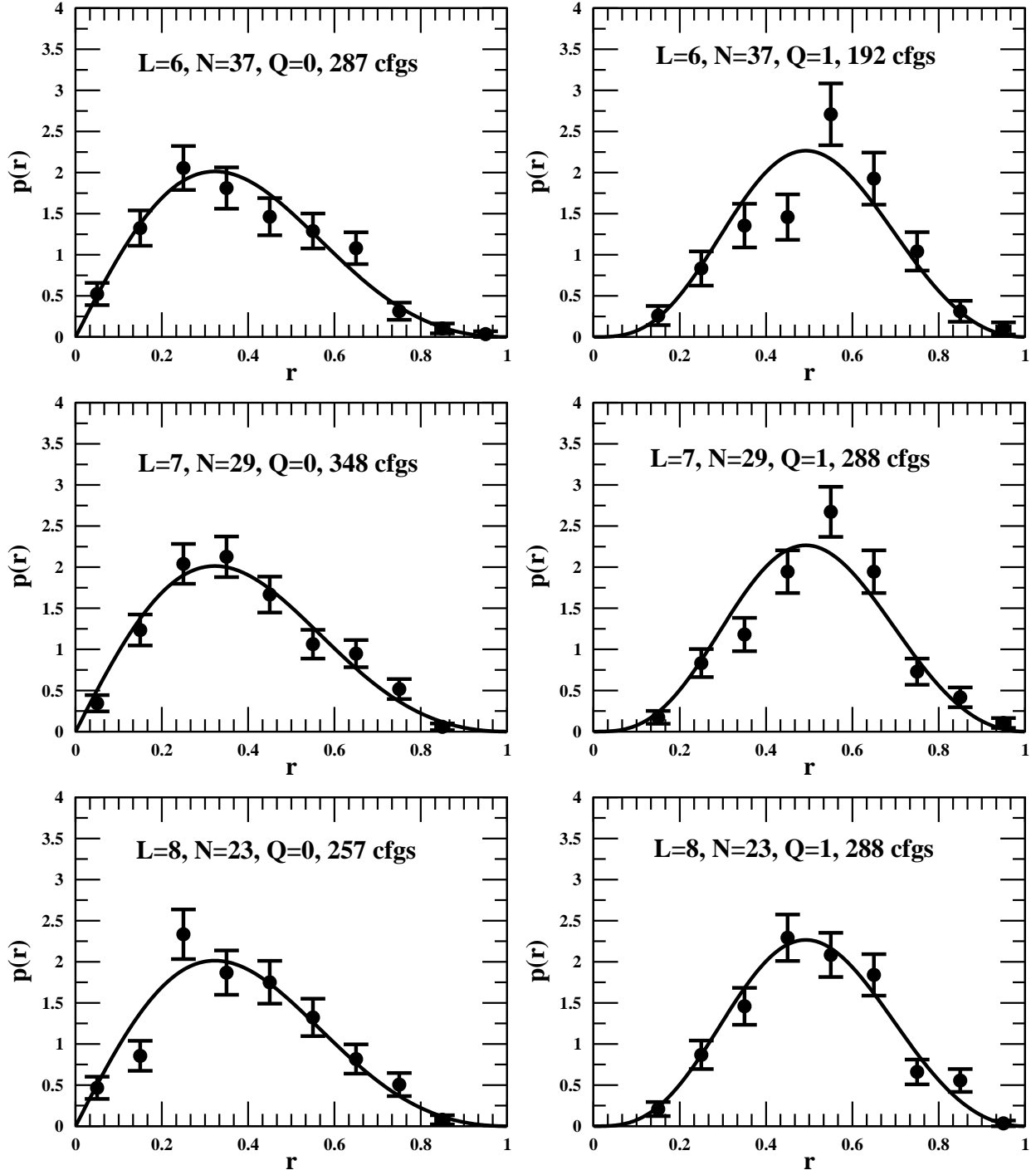}}
\caption{Evidence for agreement with chiral random matrix theory 
in the $Q=0$ and $Q=1$ topological sectors at a fixed $b=0.350$.
\label{toprmt}}
\end{figure}

\begin{figure}[ht]
\epsfxsize = \textwidth
\centerline{\epsfbox{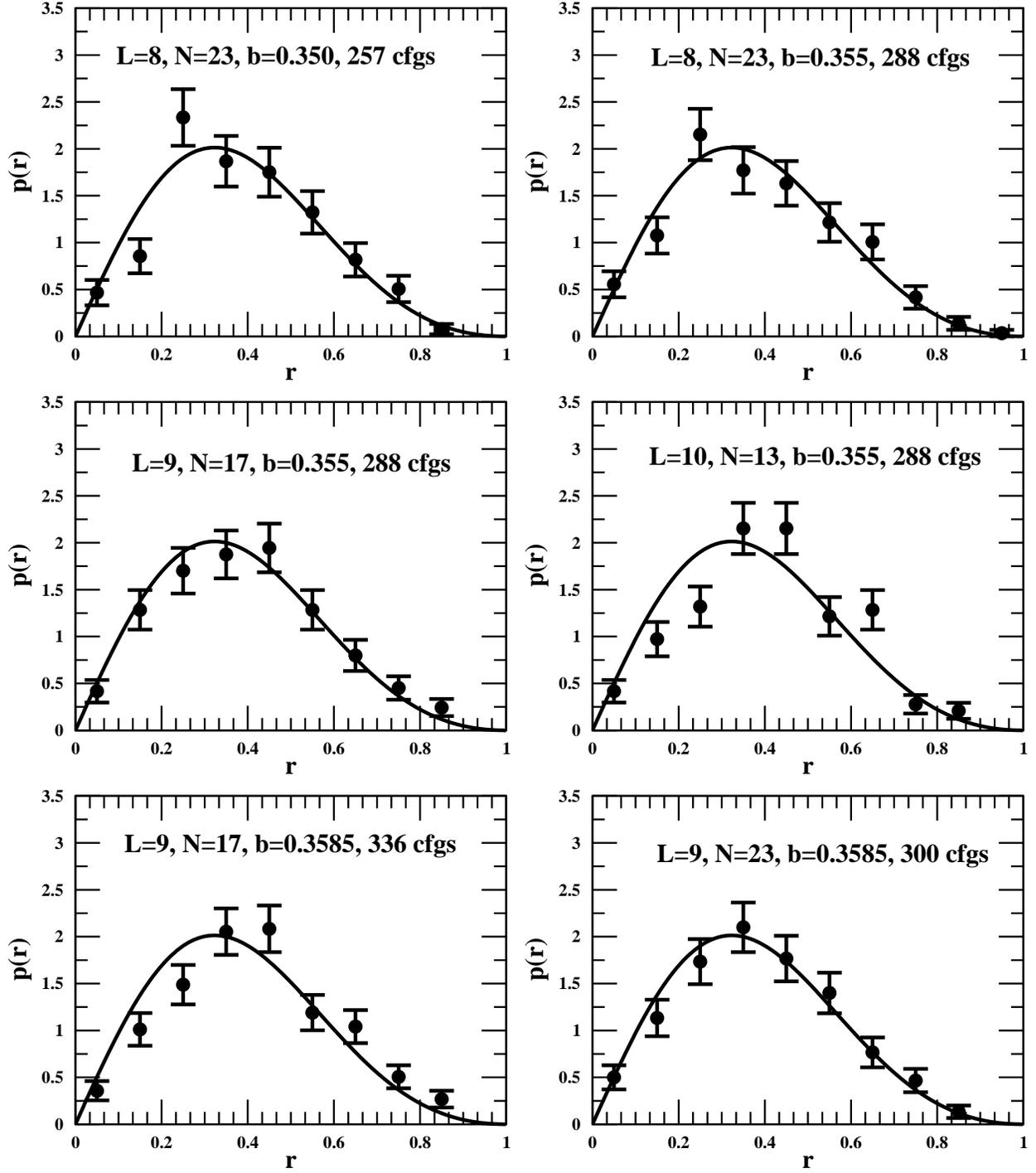}}
\caption{Evidence for agreement with chiral random matrix theory 
as one approaches the continuum limit.
\label{rmtcont}}
\end{figure}

Fig.~\ref{d4rmt} shows the approach to chiral random matrix theory as one
approaches the large $N$ limit at a fixed $L$ and $b$. One sees a clear
agreement with chiral random matrix theory in the distribution of the
ratio of the two lowest eigenvalues for $N\ge 23$.
All configurations
are in the zero topological sector.  
The universal distribution of $p(r)$ is different in the different topological
sectors and Fig.~\ref{toprmt} shows that one finds agreement with chiral
random matrix theory in the
$Q=1$ topological sector when there is agreement in the $Q=0$ topological
sector.
Finally, Fig.~\ref{rmtcont} shows that
one also finds agreement with chiral random matrix theory as
one goes toward the continuum limit.

\begin{figure}[ht]
\epsfxsize = \textwidth
\centerline{\epsfbox{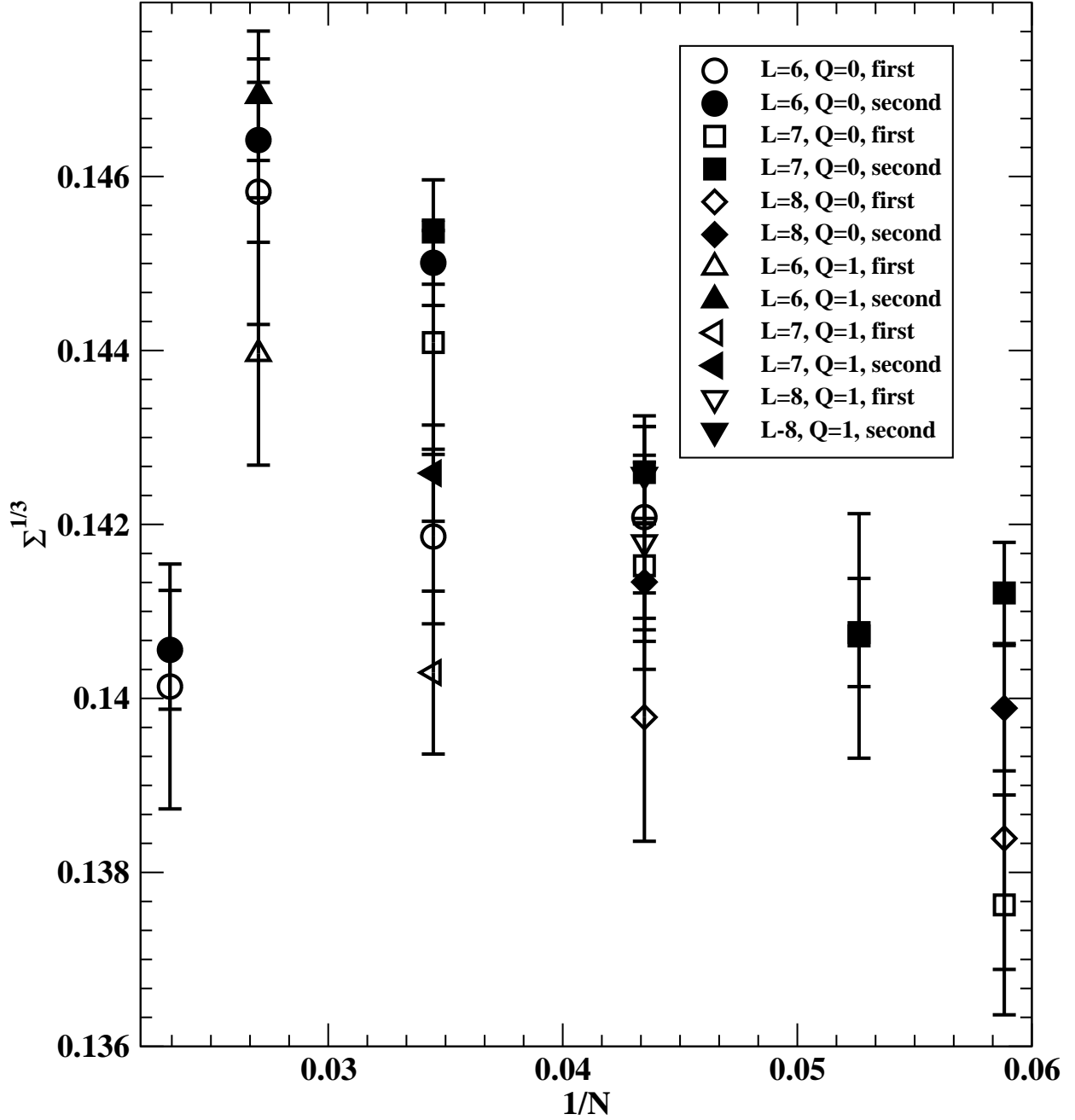}}
\caption{
Estimate of the chiral condensate in the large $N$ limit of QCD.
\label{sigma4d}}
\end{figure}

The first and second eigenvalues are used to obtain the chiral condensate,
$\Sigma$, by setting the average of $z_i=\lambda_i \Sigma_i N L^4;\ \ 
i=1,2$ equal to the ones dictated by chiral random matrix theory.
One should find $\Sigma_1=\Sigma_2$ for values of $b$, $L$ and $N$
where $p(r)$ agreed with the universal distribution given by chiral
random matrix theory. Fig.~\ref{sigma4d} shows the results for
$\Sigma_1$ and $\Sigma_2$ from all our numerical simulations at $b=0.350$ where
we found $p(r)$ to agree with chiral random matrix theory.
First of all we find that $\Sigma_1=\Sigma_2$ for all the data shown in
Fig.~\ref{sigma4d}. Furthermore, we do not see any finite volume effects
at $b=0.350$ indicating that reduction works properly for fermionic
observables. The finite $N$ effects are relatively small over the
range of $N$ shown since
the $y$-axis in Fig.~\ref{sigma4d} shows a deviation of
only $5\%$. We can therefore
estimate $\Sigma^{1/3}(b=0.350)=0.142\pm 0.006$.

Estimation of the chiral condensate using chiral random matrix theory has
proved to be far superior to conventional techniques~\cite{nn2}. 
The main problem that plagues the conventional approach is the quadratic
divergence in $\langle\bar\psi\psi\rangle$. Therefore, large subtractions are needed
in order to extract the chiral condensate. This can remedied by considering
derivatives with respect to $m^2$ of the chiral condensate. This
ameliorates the problem but the dependence on small quark masses was
still strong and this makes it difficult to 
estimate the chiral condensate using extrapolation~\cite{nn2}.
In addition to the need for subtractions, one can only obtain a stochastic
estimate of this quantity and it is necessary for the random source to have
good overlap with the low lying eigenvalues. On the other hand, numerical
evaluation 
using Ritz functional techniques
results in estimates of low lying eigenvalues with errors that are
less than a tenth
of a percent.

If we interpret the $0$c to $1$c transition as a finite temperature
phase transition, then one would expect chiral symmetry to be 
restored in the $1$c phase. But, chiral random matrix theory arguments
would indicate that chiral symmetry would still be broken since
the $Z_N$ symmetry is broken only in one of the four directions.
Three other directions still behave as if they were infinite in extent
and one would still expect level repulsion with level spacings
that go like $\frac{1}{NL^3}$. Therefore, one would expect chiral symmetry
to be broken in the $1$c phase with a value that depends on the
length of the direction which is broken. Therefore, it would be
interesting to compute the chiral condensate in the $1$c phase.
Similar arguments would also say that chiral symmetry is broken
in the $2$c and $3$c phase. Chiral symmetry is expected to be
restored only in the $4$c phase. 
A study of chiral symmetry breaking in these other phases would
be interesting on their own right for a full understanding of
the associated continuum theories.

\section{Conclusions and future work}

Phase transitions as a function of scale are ubiquitous in large $N$ QCD.
Continuum reduction~\cite{nn1} says that large $N$ gauge theory on an $l^d$ torus
undergoes a phase transition when $l=l_c$ and large $N$ QCD 
is properly reproduced
as long as $l>l_c$. This can used to numerically
solve large $N$ QCD by working on small lattices. Furthermore, one can
work in the quenched approximation on the lattice. It is important to
take the $N\rightarrow\infty$ limit before one takes any other limit.
Then, one finds that there are no finite volumes effects and it is
possible to obtain the low energy parameters of the chiral Lagrangian
in the large $N$ limit of QCD~\cite{nn2}. 
Chiral random matrix theory~\cite{rmt}
proved to be a useful tool in the extraction of the chiral condensate
since it describes the behavior of small eigenvalues for large but 
finite number of degrees of freedom.

The next step is the computation of the pion decay constant and
work is close to completion~\cite{nn3}. This will establish the
validity of momentum quenching in the computation of meson
propagators. A study of current correlators will be useful in
the extraction of the mass of the $\rho$.

On the technical side, the Wilson Dirac operator has a gap in its
spectrum at values of $M$ used in the overlap Dirac operator.
This is due to the lattice phase transition present in the large $N$
limit of QCD and is a consequence of the gap in the eigenvalue 
distribution of the single plaquette. Therefore, gauge fields come
in disconnected subspaces. In addition, the gap in $H_w(M)$ results
in a significant reduction in the computational cost of the action
of the overlap Dirac operator on a vector.

The $1$c to $4$c phases discussed in section 3 have possible
implications for certain string theories~\cite{string} and it
would be useful to compute the chiral condensate in all these
phases.

\section*{Acknowledgments}
R. N. acknowledges partial support by the NSF under
grant number PHY-0300065 and also partial support from Jefferson 
Lab. The Thomas Jefferson National Accelerator Facility
(Jefferson Lab) is operated by the Southeastern Universities Research
Association (SURA) under DOE contract DE-AC05-84ER40150.
H. N. acknowledges partial support 
by the DOE under grant number 
DE-FG02-01ER41165.
A significant part of the results presented here were obtained in
collaboration with Joe Kiskis. We thank the organizers of the
workshop for providing a stimulating environment.

\end{document}